\documentclass[12pt,a4paper]{article}
\usepackage{graphicx,amssymb,bm,latexsym,amsmath,braket,wrapfig,cite,color}
\usepackage{float}
\makeatletter
\@addtoreset{equation}{section}

\makeatother

\newcommand{\bea}{\begin{eqnarray}}
\newcommand{\eea}{\end{eqnarray}}
\newcommand{\vs}[1]{\vspace{#1 mm}}

\newcommand{\til}{\widetilde}

\newcommand{\p}[1]{(\ref{#1})}

\begin{document}
\begin{center}
{\fontsize{20}{0pt}\selectfont\bf
Quantum improved charged black holes } \\

\end{center}

\medskip
\renewcommand{\thefootnote}{\fnsymbol{footnote}}

\begin{center}
Akihiro Ishibashi${}^{1,2,} $\footnote{e-mail: akihiro@phys.kindai.ac.jp},
 Nobuyoshi Ohta$^{3,2,}$\footnote{e-mail: ohtan@ncu.edu.tw},
and
Daiki Yamaguchi$^{1,}$\footnote{e-mail: daichanqg@gmail.com} \\

\medskip

\emph{${}^1$Department of Physics, Kindai University, Higashi-Osaka, Osaka 577-8502, Japan
\\
\medskip
${}^2$Research Institute for Science and Technology, \\
Kindai University, Higashi-Osaka, Osaka 577-8502, Japan
\\
and
\\
${}^3$Department of Physics, National Central University, Zhongli, Taoyuan 320317,
Taiwan
}
\end{center}
\medskip

\begin{center}
{\bf Abstract}
\end{center}
We consider quantum effects of gravitational and electromagnetic fields in spherically symmetric black hole
spacetimes in the asymptotic safety scenario. Introducing both the running gravitational and electromagnetic
couplings from the renormalization group equations and applying a physically sensible scale identification
scheme based on the Kretschmann scalar, we construct a quantum mechanically corrected, or “quantum improved''
Reissner-Nordstrom metric. We study the global structure of the quantum improved geometry and show,
in particular, that the central singularity is resolved, being generally replaced with a regular Minkowski-core,
where the curvature tensor vanishes. Exploring cases with more general scale identifications, we further find
that the space of quantum improved geometries is divided into two regions: one for geometries with a regular
Minkowski-core and the other for those with a weak singularity at the center. At the boundary of the two regions,
the geometry has either Minkowski-, de Sitter-, or anti-de Sitter(AdS)-core.

\renewcommand{\thefootnote}{\arabic{footnote}}
\setcounter{footnote}{0}

\section{Introduction}
The occurrence of a spacetime singularity under complete gravitational collapse is a robust prediction of
general relativity, as the singularity theorem asserts~\cite{HE73}. General relativity is therefore not
a complete theory of gravity as it breaks down at spacetime singularities. The resolution of spacetime
singularities is one of the main purposes of constructing quantum theory of gravity.

The asymptotic safety scenario is one of such attempts to formulate a consistent quantum theory of gravity,
based on the renormalization group flow that controls coupling constants~\cite{R98}.
For reviews of this approach and related references, see \cite{Percacci,RS2019}. As an application to
black holes, Bonanno and Reuter~\cite{BR00} have studied possible consequences of the asymptotic safe gravity
on the Schwarzschild metric, in which the ordinary Newton constant in the classical metric component
is replaced with the running gravitational constant obtained from the renormalization group equation.
Such a quantum mechanically corrected black hole is referred to as the ``renormalization group improved'' or
``quantum improved'' black hole. Since the running Newton constant acquires a certain scale dependence,
the global structure of the quantum improved Schwarzschild black hole becomes quite different from its classical
counterpart, admitting an inner (Cauchy) horizon besides the black hole event horizon. It was shown,
in particular, that the central curvature singularity of the classical Schwarzschild metric is resolved to
be a regular center, neighborhood of which is well described by the de Sitter metric. In this regard, the global
structure of the quantum improved Schwarzschild black hole is similar to that of the Bardeen black
hole~\cite{Bardeen}.

Although the appearance of the Cauchy horizon raises new issues such as the breakdown of the strong cosmic
censorship, the resolution of a classical singularity is a fascinating aspect anticipated for the asymptotically
safe gravity to be a successful candidate of quantum theory of gravity. Subsequently further analyses of quantum
improved black holes have been made, and a number of basic properties, such as the horizon structure,
the thermodynamic laws, and Hawking temperature, have been studied
in detail~\cite{BR2006,RT11,FLR12,KS141,KS142,BKP2017,PS18,Pla2019,BCP2019,Plat2020}.
For instance, the analysis was generalized to include the rotation (angular momentum) parameter and a cosmological
constant (either positive or negative), with the latter replaced with the running cosmological constant~\cite{PS18}.
It was found that the curvature singularity of the quantum improved Kerr-(A)dS geometry becomes less divergent
compared to the classical counterpart but is not fully removed.

So far, most of the analyses of quantum improved black holes have focused on quantum effects of pure gravity,
rather than combined effects of gravity and other fundamental forces. However, when considering quantum effects
around a singularity, it is natural to ask whether fundamental forces other than gravity play any significant role.
It is far from obvious if, among all, gravity is the most dominant force to shape the structure of singularities.
The contribution of matter fields to the evolutions of the cosmological constant and Newton constant has been
studied in \cite{DEP,MPR,EL,BPS}.
Even within classical framework, including, for example, electromagnetic interaction drastically modifies
the global structure of the otherwise vacuum Schwarzschild solution by rendering the central singularity timelike
hidden inside the Cauchy horizon, as the classical Reissner-Nordstrom metric exhibits. In view of this, it is of
considerable interest to explore if and how quantum effects of the electromagnetic field together with gravity
affect the spacetime structure especially near singularities in the context of the asymptotic
safety~\cite{HR2011,CE2017,EV18}.
For other related approach to the resolution of singularities, see \cite{LS,BE}, and also, e.g., \cite{CK21,JLQ21}.

In this paper, we consider the problem of singularity resolution by examining yet another quantum improved
black hole including the running gravitational as well as $U(1)$ gauge couplings: that is, the ``quantum improved
Reissner-Nordstrom metric,'' as a first step toward understanding fundamental interactions all together
in the asymptotic safety scenario. For this purpose, we need to address the following two issues: First,
when taking into account quantum effects of the electromagnetic field, one has to deal with Landau poles,
which involve the (logarithmic) divergence of the running $U(1)$ gauge coupling at a finite momentum scale.
To handle this problem we employ the method developed by Eichhorn and Versteegen~\cite{EV18}, which exploits
the ultraviolet completion for $U(1)$ gauge theory induced from asymptotic safe quantum gravity. Second, one
has to determine how to identify the cutoff scale with physical distance.
See \cite{BEG} for discussions on issues with such identification.
So far, several different schemes for the scale identification have been proposed.
In this paper we employ the scheme based on the Kretschmann scalar considered by Pawlowski and Stock~\cite{PS18}.
This has the advantage that it is a diffeomorphism invariant scheme and is also easy to compare the case
of classical vacuum (e.g., Schwarzschild) solution with that of non-vacuum (e.g., Reissner-Nordstrom) solution.
With these strategies at our hand, we will construct a quantum improved Reissner-Nordstrom metric.
We show that the singularity of our quantum improved Reissner-Nordstrom black hole is in general resolved
so that in the neighborhood of the center the curvature tensor vanishes. Namely, the quantum improved
Reissner-Norstrom black hole generally admits the ``Minkowski-core.''
For comparison, we also consider the type of scale identification in which the cutoff momentum is given by
the general inverse power law and show that in the space of quantum improved Reissner-Nordstrom solutions,
the geometries with either de Sitter(dS)-core or anti-de Sitter(AdS)-core appear as the boundary between the region
for regular solutions and that for weakly singular solutions.

This paper is organized as follows. In section \ref{sec:2}, we briefly review the quantum improved
Schwarzschild spacetime studied in Ref.~\cite{BR00}, which is regular at the center with a dS-core.
We also discuss how the running Newton constant depends on the cutoff momentum scale $k$, and how to identify
the scale $k$ with the inverse of the distance $r$ from the center of spherical symmetry. In section~\ref{sec:3},
we study possible combined effects of the $U(1)$ gauge field together with quantum gravity on
the Reissner-Nordstrom metric. We derive both the running gravitational and $U(1)$ gauge couplings from
the renormalization group equations. We find that our quantum improved Reissner-Nordstrom spacetime in general
possesses a regular core at the center.
As the final topic of section~\ref{sec:3}, we consider the scale identification scheme of the inverse power
radial dependence and study the condition for quantum improved solutions with regular center and that
for weakly singular solutions. Section~\ref{sec:4} is devoted to summary and discussions.

\section{Quantum improved Schwarzschild black hole}
\label{sec:2}

In this section we briefly review the result of Ref.~\cite{BR00}. The Schwarzschild spacetime is
the unique spherically symmetric vacuum solution of the Einstein equation. The Schwarzschild metric is written as
\bea
ds^2=-f(r)dt^2+\frac{1}{f(r)}dr^2+r^2 (d\theta^2 + \sin^2\theta d\phi^2) \,,
\label{def:metric}
\eea
where the metric function is given by
\bea
f(r)=1-\frac{2G_{0}M}{r} \,,
\label{2.1}
\label{def:Schwa}
\eea
with the mass parameter $M$ and the Newton constant $G_{0}$. Hereafter, the metric function $f(r)$ is called
the lapse function.

In Ref.~\cite{BR00}, quantum gravitational effects in the Schwarzschild spacetime have been studied, by using
the renormalization group method and introducing the running Newton constant. The first step is to derive
the scale dependent Newton constant $G(k)$ by considering the evolution of scale-dependent effective
gravitational action by means of an exact renormalization group equation. The evolution of the dimensionless
Newton constant ${\til G}(k) := k^2 G(k)$ is governed by
\bea
k\frac{d {\til G}}{dk}= \beta( {\til G}(k)) \,,
\eea
where the beta function $\beta({\til G})$ is given by eq.~(2.13) of Ref.~\cite{BR00}.
This has two (i.e. IR and UV) fixed points. By inspecting the UV attractive fixed point,
the dimensionful Newton constant $G(k)$ is obtained, with the identification $G_0 = G(k=0)$, as
\bea
G(k)=\frac{G_{0}}{1+\omega G_{0} k^2} \,,
\label{eq:Gk}
\eea
where $\omega$ is the constant which can essentially be identified with the inverse of the UV fixed point
(see discussion around eqs.~(2.21)-(2.25) of Ref.~\cite{BR00}).

The next step is to identify the cutoff scale $k$ with the radial distance $r$ from the center of
the spherical symmetry. Such a scale identification is schematically written as
\bea
k=\frac{\xi}{d(r)} \,,
\label{2.2}
\eea
where $d(r)$ denotes an appropriate function of $r$ and $\xi$ a numerical constant.  So far, several different
choices of $d(r)$ have been proposed. For example, one can choose $d(r)$ as the proper distance from an observer
to the center, along a straight (non-null) radial path. In this case, one obtains $k(r) \propto 1/r^{3/2}$
in the UV regime.

The idea of constructing the quantum improved Schwarzschild metric is to obtain the position-dependent
Newton constant $G(r)=G(k(r))$ by inserting a certain choice of the scale identification \p{2.2} into
the running Newton constant $G(k)$, and then to replace the Newton constant $G_{0}$ in the classical
lapse function \p{2.1} by the position-dependent counterpart $G(r)$. By doing so, one obtains the quantum
improved lapse function
\bea
f(r)=1-\frac{2M}{r}G(r) \,.
\eea
Note that as will be described below, instead of replacing the classical gravitational constant $G_0$
in the metric function, one can also take
the replacement of $G_0$ in the Einstein equations or $G_0$ in the Einstein-Hilbert action.

For the choice of $d(r)$ as the proper radial distance, one finds that
$d(r)=r[1+O(1/r)]$ at large $r$ and $d(r) = r^{3/2}\sqrt{2/(9G_0M)} + O(r^{5/2})$ near the center $r=0$
and therefore that the position-dependent Newton constant becomes
$G(r) \simeq G_0 - \omega \xi^2 G_0^2/r^2 + O(1/r^3)$ at large $r$ and
$G(r) \simeq 2r^3/9{\omega} \xi^2 G_0 M$ near $r=0$. Hence, in particular, the quantum improved lapse
function behaves near the center as
\bea
f(r)=1-\frac{4}{9 {\omega}\xi^2 G_0}r^2+\mathcal{O}(r^3) \,.
\eea
Thus, the quantum improved lapse function is regular near the center and describes the de Sitter geometry
with the curvature length $l=(3/2)\xi \sqrt{\omega G_0}$. Such a near-center region is called the ``de Sitter core.''
Note that according to Ref.~\cite{BR00}, for some cases of interest, by comparing the behavior
of $G(r)$ at large $r$ with the results of the perturbative quantization of the Einstein gravity,
the value of $\omega \xi^2$ can be determined as $\omega \xi^2 =118/(15 \pi)$.

The choice of $d(r)$ determines the $r$-dependence of the momentum scale $k$. However, such a scale identification
is not unique, in particular, for pure gravity systems. So far, several different scale identifications
have been considered.
In the above example~\cite{BR00}, $d(r)$ is taken as the proper radial distance. Another diffeomorphism
invariant procedure is to express $d(r)$ in terms of the curvature scalars, such as the scalar curvature,
or the Kretschmann scalar of either the classical geometry or the corresponding quantum improved geometry.
We list some examples of $d(r)$ relevant to our subsequent analyses and their behavior in the UV regime
in Table~\ref{table1}. For more detailed list, see Ref.~\cite{PS18}.
See also Ref.~\cite{Held21} for discussion on the use of various curvature invariants in quantum improvements of spherically
symmetric and axially-symmetric black hole spacetimes and their coordinate (in)dependence.
\begin{table}[htb]
\begin{center}
\caption{Behavior of $d(r)$ in the short distance limit. Numerical factors are dropped.}
\vspace{2mm}
\begin{tabular}{|c|c|c|c|} \hline
    & Kretschmann & Radial path& Geodesic path \\ \hline
    Schwarzschild (Classical) &  $r^{3/2}$ & $r^{3/2}$ & $r^{3/2}$ \\ \hline
    Schwarzschild (Quantum) & $r^{3/4}$ & $r$ & $r^{3/4}$ \\ \hline
\end{tabular}\label{table1}
\end{center}
\end{table}

Let us briefly summarize the prescription for constructing quantum improved geometry.
For more details, see, e.g., \cite{BR00,PS18,Plat2020,RW04}.
\begin{itemize}
\item[{\bf Step 1.}] Derive the scale-dependent coupling constants from exact renormalization group (RG) equations:
 e.g.,
$k \dfrac{d(k^2G)}{dk}=\beta (k^2G)$ for the running gravitational constant $G(k)$.

\item[{\bf Step 2.}] Find a cutoff scale identification $k=k(x)$ and convert the scale dependence of
the couplings into the position dependence: e.g., $G(k) \rightarrow G(x):=G(k(x))$.
For static spherically symmetric geometries, it is natural to assume that $k$ be a function only of
the area radius $r$.
As briefly shown in Table~\ref{table1}, there have been several different schemes for scale identifications,
which may roughly be classified into two types:

\begin{enumerate}
\item[{\bf (i)}] Based on curves toward a singular point as the UV limit: e.g., $k \propto 1/d(r)$ with
$d(r)$ being the proper radial distance.
\item[{\bf (ii)}] Based on invariant curvature scalars: e.g., such as $R$, $R^\alpha{}_\beta R^\beta{}_\alpha$,
or $K=R_{\mu \nu \alpha \beta}R^{\mu \nu \alpha \beta}$ with $k \propto K^{1/4}$.
\end{enumerate}
One may use the relevant geometric quantities above with respect either to the classical geometry or to
the quantum improved geometry~\cite{PS18}.

\item[{\bf Step 3.}]  Perform quantum improvement. There are three approaches:
\begin{enumerate}
\item[{\bf (i)}] {\bf Improved solutions.} One considers a classical solution and replaces the couplings appearing
in the classical solution with the position dependent ones: e.g., $G_0$ in the classical lapse function $f(r)$
is replaced with $G(r)$.

\item[{\bf (ii)}] {\bf Improved equations of motion.} One replaces the couplings with the position dependent
ones at the level of equations of motion:
e.g., one needs to solve the improved Einstein equations, $G_{\mu \nu} = 8\pi G(x)T_{\mu \nu}
 - \Lambda(x) g_{\mu \nu}$,
where $\Lambda(x)$ is the running cosmological constant and where the running of matter couplings are ignored.

\item[{\bf (iii)}] {\bf Improved action.} One replaces the couplings with the position dependent ones at
the level of action, derives the modified field equations from the quantum improved action, and then solves
the modified field equations. For instance, the quantum improved Einstein-Hilbert action is given by
the modified Lagrangian,
$L = \dfrac{1}{16 \pi G(x)}\left\{R-2 \Lambda(x) \right\}$.
\end{enumerate}

These three approaches make, in principle, different predictions. For instance, the equations of motion derived
from the quantum improved Einstein-Hilbert action can contain the derivatives of $G(r)$, while the quantum
improved Einstein equation does not.
The improvement at the level of action is expected to be most natural and have highest predictive power.
The improvement at the level of
solutions is the simplest approach but is thought to be applicable only when the quantum improved metric
is not drastically different from the classical counterpart.
\end{itemize}

In the next section, we will consider quantum improvement of the Reissner-Nordstrom geometry.
With the electromagnetic coupling $e_0$ in addition to $G_0$, we follow Steps 1 and 2. (We set $\Lambda =0$.)
Then, with two position dependent couplings $e(r), G(r)$ at hand, the hardest part of the job would be
to perform Step 3 if one takes the approach of either (ii) the improved Einstein equations, or
of (iii) the improved Einstein-Hilbert-Maxwell action.
In this paper, we shall take (i) the improved solution approach, following \cite{BR00}.

\section{Quantum improved Reissner-Nordstrom black hole}
\label{sec:3}

In this section, we study quantum effects of both gravity and electromagnetic fields on the static
spherically symmetric charged black hole geometry. For this purpose, we take, as our classical background,
the Reissner-Nordstrom metric of the form (\ref{def:metric}) with the classical lapse function
\bea
f(r)=1-\frac{2G_0M}{r} +\frac{G_0 e_0^2}{r^2} \,,
\label{3.15}
\label{metric:RN}
\eea
where $e_0$ is the classical $U(1)$ coupling.
We start with deriving the running Newton constant $G(k)$ and the running $U(1)$ gauge coupling $e(k)$ from
the functional renormalization group equations. Each of these two running couplings acquires a certain
dependence on the radial distance $r$ through a scale identification with the momentum cutoff $k$.
As a consequence, we obtain a quantum mechanically improved Reissner-Nordstrom metric. We study the global
structure --- in particular, the central singularity --- of the quantum mechanically improved
Reissner-Nordstrom black hole.

\subsection{Running couplings}

When considering quantum effects of electromagnetic interaction, one must deal with Landau poles.
For this purpose we employ the method recently proposed by Eichhorn and Versteegen~\cite{EV18},
in which the UV completion for $U(1)$ gauge theory is successfully achieved
by using the functional renormalization group in $U(1)$-coupled gravity theory. Although the running of
the cosmological constant $\Lambda$ is also considered in Ref.~\cite{EV18},
in this paper, we restrict our attention to the Newton and $U(1)$ gauge couplings.
Then, our beta functions for running gravitational and $U(1)$ gauge couplings are given as follows (see also~\cite{HR2011}):
\bea
k\frac{d \til{G}}{dk} &=& 2\til{G} \left( 1 - \frac{1}{4\pi \til\alpha} \til{G} \right) \,,
\label{3.1}
\\
k \frac{de}{dk} &=&  \frac{1}{4\pi} e\left( \frac{b}{ 4\pi} e^2 - \til{G}\right) \,,
\label{3.2}
\label{rge:e}
\eea
where $\til\alpha, b$ are the parameters which specify the fixed points $\til{G}_{*}, e_{*}$ as
\bea
\til{G}_{*}
= 4\pi \til\alpha \, , \quad e_{*}^2 = (4\pi)^2\frac{\til{\alpha}}{b} \,.
\label{3.4}
\eea
Note that for the Schwarzschild background case, one may identify $\til \alpha$ with $\omega$ in eq.~(\ref{eq:Gk}) as
\bea
\label{def:tilalpha}
 \omega = \frac{1}{4 \pi \til\alpha}\,.
\eea

One may specifically determine the value of $b$ in terms of the $U(1)$ charges $(Q_{{\rm D}i},Q_{{\rm S}i})$
of the matter (Dirac and scalar) fields:
\bea
b=\frac{4}{3} \sum_i (Q_{{\rm D}i})^2+\frac{1}{3} \sum_i (Q_{{\rm S}i})^2 \,,
\eea
where the subscripts ${\rm D}$ and ${\rm S}$ stand for Dirac and for charged scalar fields, respectively,
and $i$ denotes the kinds of the fields.
In the context of the Standard Model in particle physics, the fixed point value of the $U(1)$ gauge coupling is
given with the value $b= 41/6$ (see eq.~(4.1) in Ref.~\cite{EV18}).
If one chooses
\bea
N_{\rm V} = 1+2N_{\rm D}-\frac{N_{\rm S}}{2},
\label{vac}
\eea
with $N_{\rm V}, N_{\rm D}, N_{\rm S}$
being the numbers of vector, Dirac and scalar fields, respectively,
then the cosmological constant may be set to zero consistently with the renormalization group equation
(namely it is a fixed point).
Though this relation is not satisfied in the Standard Model in which $N_{\rm V}=12, N_{\rm D}=45/2$ and $N_{\rm S}=4$,
there may be additional particles like dark matter
and we can expect that this may be satisfied in particle physics models beyond the Standard Model.

We will find that the singularity is resolved only for vanishing cosmological constant,
and then the condition~\p{vac} could be even regarded as an interesting criterion to select theories.
The difference of the rhs and lhs of Eq.~\p{vac} for the Standard Model is 32, which requires 64
additional scalar fields as dark matter.
Another example that is closer to satisfying the condition~\p{vac} is the minimal supersymmetric standard model
where we have additional supersymmetric partners, $\Delta N_{\rm D}=(12+4)/2$ and $\Delta N_{\rm S}=94$
(we must include a second Higgs doublet in supersymmetric theory).
This is close to satisfying \p{vac} (the difference of the rhs and lhs of Eq.~\p{vac} reduces to 1) but not precisely.
We could also consider $SU(5)$ grand unified theory, where we have 24 gauge bosons, and two Higgs fields
in the representations of {\bf 24} and {\bf 5} to break $SU(5)$ gauge symmetry down to $SU(3)\times SU(2)\times U(1)$ and to give masses to quarks and leptons, respectively.
Here we have $N_{\rm V}=24, N_{\rm D}=45/2$ and $N_{\rm S}=34$, and the difference is 5. In this case,
we could consider an additional (complex) scalar multiplet in the {\bf 5} representation of $SU(5)$ for
the dark matter. With these considerations and other possibilities in mind, we assume the relation~\p{vac}
in this paper. This also simplifies our analysis.

By comparing \p{3.2} with the renormalization group equation for the $U(1)$ coupling in \cite{EV18},
we get the relation
\bea
\frac{1}{\til\alpha}=\frac{-2N_{\rm D}-N_{\rm S}+4N_{\rm V}+29}{3}.
\eea
Upon eliminating $N_{\rm V}$ by use of the relation~\p{vac}, we have
$\til\alpha = {1}/{(11+2N_{\rm D}-N_{\rm S})}$.
One may expect $\til\alpha$ to take some value much smaller than one: $\til\alpha \ll 1$.
For example, if we take $N_{\rm D}=N_{\rm S}=1$ for simplicity, then $\til \alpha = 1/12$.
For $b=41/6, N_{\rm D}=45/2$ and $N_{\rm S}=4$, we have ${\til\alpha} = {1}/{52}$,
$e_*^2 = (4\pi)^2 {\til{\alpha}}/{b}= {24\pi^2}/{533} \approx 0.444$.
However, when one is concerned with the UV regime very close to the (classical) central singularity,
it is far from obvious that one can adequately estimate model parameters by the low-energy degrees of freedom.
Rather one may anticipate some nontrivial contributions from physics beyond the Standard Model.
Taking account of such possibility as well as the developing nature of the asymptotic safety scenario,
in this paper, we assume that $\til \alpha$ takes a wider parameter range: $0< \til\alpha \lesssim 1$.

Integrating \p{3.1} from a reference scale $k_0$ to $k$ and using the relation $\tilde G(k)=k^2G(k)$,
we obtain the running Newton constant
\bea
G(k)=\frac{ 4\pi {\til \alpha} G(k_{0})}{4\pi {\til \alpha} +  (k^2-k^2_0)G(k_0)} \,.
\label{3.5}
\eea
In what follows, we assume $k_0=0$ and set $G_0=G(k_0=0)$ as in the Schwarzschild background case.
Then, our running Newton constant becomes,
\bea
G(k) = \frac{4\pi \til\alpha G_0}{4\pi \til\alpha + G_0k^2} \,.
\label{3.6}
\eea
Note that this is precisely the same as (\ref{eq:Gk}) with the identification~(\ref{def:tilalpha}).
The asymptotic behaviors of $G(k)$ at short and large distances are
\bea
G(k) \simeq \frac{4 \pi \til\alpha}{k^2} \quad (k \rightarrow \infty)\,,
\quad
G(k) \simeq G_0 \quad (k \rightarrow 0)\,.
\label{G:asymptbehave}
\eea

As for the $U(1)$ gauge coupling, by introducing the following functions,
\bea
P(k)  :=\frac{\til\alpha G_0 k}{4\pi \til\alpha + G_0k^2} \,, \quad
Q(k) := \frac{b}{(4\pi)^2 k} = \frac{\til\alpha}{e_{*}^2} \cdot \frac{1}{k} \,,
\label{3.7}
\eea
we can express \p{3.2} in the form of the Bernoulli equation,
\bea
\frac{de}{dk}+P(k)e=Q(k)e^3 \,.
\label{3.8}
\eea
Integrating \p{3.8}, we obtain the formal solution:
\bea
\frac{1}{e^2(k)}
 = \exp \left({2\! \int \! d{k}P({k})} \right) \left[C_0-2\! \int \! dkQ(k)
\exp\left({-2\! \int \! dk P(k)} \right)\right] \,,
\label{3.9}
\eea
where $C_0$ is an integration constant.
A similar formula is also obtained in~\cite{HR2011}.
To proceed further, we need to treat two cases,
$0<\til\alpha < 1$ and $\til \alpha =1$, separately.

\subsubsection{Case $\til{\alpha} \neq 1 ( \til\alpha >0 )$}
\label{subsec:ale1}

In this case, inserting \p{3.7} in the solution~(\ref{3.9}), we obtain the $U(1)$ gauge coupling
\bea
\frac{1}{e^2(k)} = C_0 \left(1+Dk^2 \right)^{\til\alpha}
+ \frac{1}{e_{*}^2} \frac{\til \alpha }{(1-\til \alpha )} (1+Dk^2)F\left(1,1-\til\alpha,2-\til\alpha;1+ Dk^2\right) \,,
\label{3.10}
\eea
where $F$ denotes the hypergeometric function, and here and hereafter we use the abbreviation
$D:= G_0/4\pi \til \alpha$ for notational brevity.
Applying the linear transformation formula of the hypergeometric function~\cite{AS65}, we can express (\ref{3.10}) as
\bea
\frac{1}{e^2(k)} = C \left(1+Dk^2 \right)^{\til\alpha}
  + \frac{1}{e_{*}^2} \cdot F\left(1, \til\alpha, 1+ \til\alpha ; \frac{1}{1+ Dk^2} \right) \,,
\label{eq:3.10b}
\eea
where we have defined a new constant $C$ involving the integration constant $C_0$ in \p{3.9}.
We will see that this is related to the magnitude of the electric charge at low energy.
%
From this expression, we can immediately read off the UV behavior $k\rightarrow \infty$ as
\bea
\label{g:UV}
\frac{1}{e^2(k)} \simeq C\cdot D^{\til\alpha} \cdot k^{2 \til\alpha} + \frac{1}{e_*^2} +O(k^{2(\til\alpha -1)}) \,.
\eea
Note that when $\til\alpha >1$, the term of $O(k^{2(\til\alpha -1)})$ becomes larger than $1/e_*^2$,
but is anyway subleading.

We can further transform the solution~(\ref{eq:3.10b}) into the form
\bea
&{}& \frac{1}{e^2(k)} = C \cdot \left(1+Dk^2 \right)^{\til\alpha}
\nonumber \\
&{}& \quad + \frac{\til\alpha }{e_{*}^2} \cdot \sum_{n=0}^\infty \frac{(\til\alpha)_n}{(n!)^2}
    \left\{ \psi(n+1)-\psi(\til\alpha+n) - \log\left( \frac{Dk^2}{1+ Dk^2} \right)  \right\}
    \left( \frac{Dk^2}{1+Dk^2}\right)^n  \,,
\eea
where $\psi$ denotes the digamma function and $(\til\alpha)_n:=\Gamma(\til\alpha +n)/\Gamma(\til\alpha)$.
From this expression, we can read off the IR behavior $k \rightarrow 0$ as
\bea
\label{g:IR}
\frac{1}{e^2(k)} \simeq C - \frac{\til\alpha}{e_{*}^2} \left[\gamma+\psi(\til\alpha) \right]
 + \frac{\til\alpha}{e_{*}^2} \cdot \log\left( \frac{1+ Dk^2}{Dk^2} \right)\,,
\eea
where $\gamma = - \psi(1)$ 
is the Euler constant.
Although the logarithmic divergence appears in the IR expression above, one can now single out the arbitrary
(integral) constant $C$  and may fix its value in terms of the classical counterpart $e_0$ of the $U(1)$ coupling.

Some comments on the constant $C$ are in order.
First note that given a finite positive value of $\til\alpha$, the second term of rhs of \p{g:IR} is
finite (as ${\til\alpha} \left[\gamma + \psi(\til\alpha)\right] = -1 + O({\til\alpha}^2)$
for small $\til\alpha$) and the third term is positive.
This third term grows for small $k$ and makes \p{g:IR} positive for sufficiently small $k$
whatever the value of the constant $C$ is. It then appears from the UV expression of (\ref{g:UV}) 
that the solution (\ref{eq:3.10b}) may change its sign for negative $C$ when varying $k$ from 0 to $\infty$.
However this should not happen because, according to the differential equation~(\ref{rge:e}), its solution stops
running when it comes close to zero before it changes the sign. The point where $1/e^2$ becomes zero
corresponds to a singularity, beyond which the solution cannot be extended.
The correct interpretation is that
\begin{enumerate}
\item[(i)] for negative $C$ in which the solution appears to change sign, the $U(1)$ coupling
diverges and the running stops there. This is what is known as Landau pole;
\item[(ii)] for positive $C$ in which the solution does not change its sign, we have the $U(1)$ coupling
for the whole range of $0<k<\infty$, and it becomes zero asymptotically;
\item[(iii)] for special critical boundary condition $C=0$, it becomes constant asymptotically.
\end{enumerate}

Let us define the $U(1)$ coupling at the energy $k=k_L \equiv 1$ GeV by
\bea
\frac{1}{e_0^2} = C(1+D k_L^2)^{\til\alpha} +\frac{1}{e_*^2}
 F\left(1, \til\alpha, 1+ \til\alpha; \frac{1}{1+ Dk_L^2} \right).
\label{e0def}
\eea
For the choice of $\til\alpha=1/52$, $e_*^2=24\pi^2/533$ and
the Newton constant $G_0=6.7\times 10^{-39}$ GeV, we find
the corresponding values between $C$ and $e_0^2$ as given in Table~\ref{table2}.
\vspace{-1mm}
\begin{table}[htb]
\begin{center}
\caption{The corresponding values of $C$ and $e_0^2$ for $\til\alpha=1/52$ and $e_*^2=24\pi^2/533$.}
\vspace{2mm}
\begin{tabular}{|l|l|l|l|l|l|l|l|} \hline
$C$     &  $-0.7$ & $-0.4$ & $-0.1$ & 0     & 0.1   & 0.4   & 0.7    \\ \hline
$e_0^2$ &  0.189  & 0.179  & 0.170  & 0.167 & 0.164 & 0.156 & 0.149 \\ \hline
\end{tabular}\label{table2}
\end{center}
\end{table}
\\
For comparison, we also show the correspondence for $\til\alpha=2/3$ and $e_*^2=24\pi^2/533$
in Table~\ref{table3}. This is to be compared to the standard value $e^2=1/137=0.00730$.
\begin{table}[htb]
\begin{center}
\caption{The corresponding values of $C$ and $e_0^2$ for $\til\alpha=2/3$ and $e_*^2=24\pi^2/533$.}
\vspace{2mm}
\begin{tabular}{|l|l|l|l|l|l|l|l|} \hline
$C$     &  $-0.7$ & $-0.4$ & $-0.1$ & 0     & 0.1   & 0.4   & 0.7    \\ \hline
$e_0^2$ &  0.00738  & 0.00737  & 0.00735  & 0.00734 & 0.00734 & 0.00732 & 0.00731 \\ \hline
\end{tabular}\label{table3}
\end{center}
\end{table}

\begin{figure}[htb]
\begin{center}
\includegraphics[width=90mm]{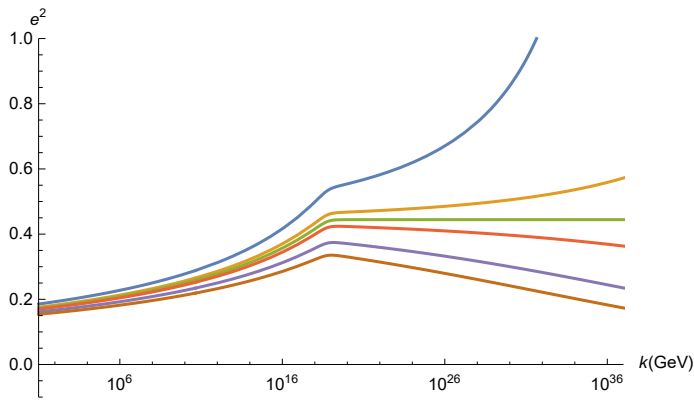}
\end{center}
\caption{
The behaviors of $e^2(k)$ for $\til\alpha=1/52$ and $e_*^2=24\pi^2/533$. The lines correspond to
$e_0^2=0.179, 0.170, 0.167, 0.164,0.156$ and $0.149$ from top to botom.}
\label{fig:U(1)c1}
\end{figure}

Figure~\ref{fig:U(1)c1} is the plot of the $U(1)$ running coupling (\ref{3.10}) for the case of $\til\alpha=1/52$
and $e_*^2=24\pi^2/533$,
in which we can see that the $U(1)$ coupling diverges for larger values than that corresponding to $C=0\,(e_0^2=0.167)$
whereas it converges to zero asymptotically for the values below that. For the critical value of $e_0^2=0.167$,
the coupling goes to the finite value (see the green curve in Fig.~\ref{fig:U(1)c1}).

The behavior of the $U(1)$ coupling for $\til\alpha=2/3$ is more or less the same as the case for $\til\alpha=1/52$, but it diverges at lower energies
than those for $\til\alpha=1/52$ when $C<0$.

\subsubsection{Case $\til\alpha=1$}

Integrating \p{3.2} with the parameter value $\til{\alpha}=1$ (hence $b=(4\pi)^2/e_*^2$), we find
the general solution:
\bea
\frac{1}{e^2(k)}= \left( 1+ \frac{G_0}{4 \pi} k^2 \right)
\left[ C_0 - \frac{1}{e_{*}^2} \cdot \log\left(\frac{G_0 k^2}{4\pi + G_0 k^2} \right)  \right] \,,
\label{4.1}
\eea
where again $C_0$ denotes an integration constant. Noting that $\gamma+ \psi(\til\alpha =1)=0$, one can see
that the IR behavior of $e^2(k)$ is the same as (\ref{g:IR}) with $C$ replaced by $C_0$
and $\til\alpha$ set to $\til \alpha=1$.
We again define the $U(1)$ coupling at the energy scale $k=k_L\equiv 1$ GeV by
\bea
\frac{1}{e_0^2}= \left( 1+ \frac{G_0}{4 \pi} k_L^2 \right)
\left[ C_0 - \frac{1}{e_{*}^2} \cdot \log\left(\frac{G_0 k_L^2}{4\pi + G_0 k_L^2} \right)  \right] \,,
\eea

The UV behavior of the $U(1)$ coupling for positive $C_0$ is
\bea
\frac{1}{e^2(k)}&\simeq & C_0 \frac{G_0 k^2}{4 \pi} \,.
\label{4.2}
\eea
For negative $C_0$, it diverges at some energy and the theory is not well defined beyond this energy.
Note that for $\til\alpha=1$, there is no critical case in contrast to $\til\alpha\neq 1$.

\begin{figure}[htb]
\begin{center}
\begin{minipage}[cbt]{.4\linewidth}
\includegraphics[width=70mm]{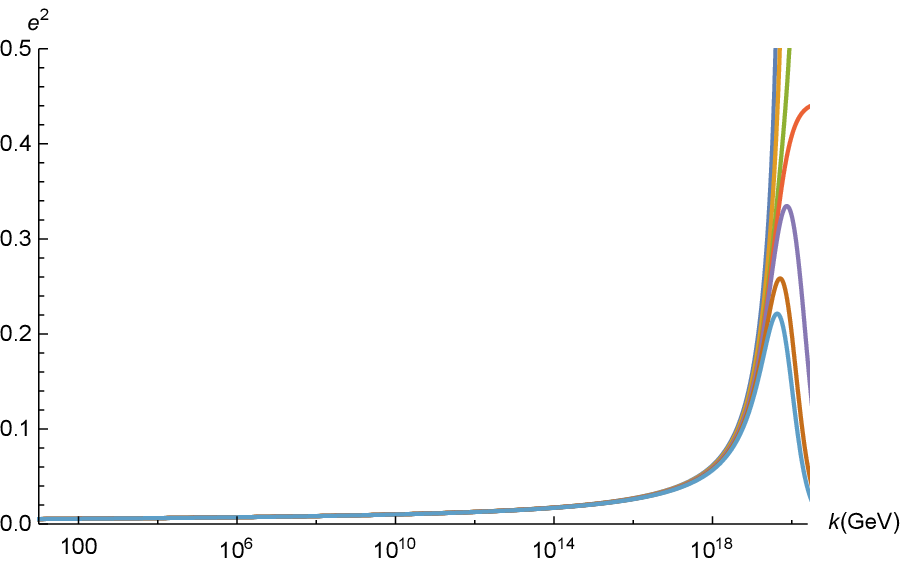}
\end{minipage}
\hspace{1cm}
\begin{minipage}[cbt]{.4\linewidth}
\includegraphics[width=70mm]{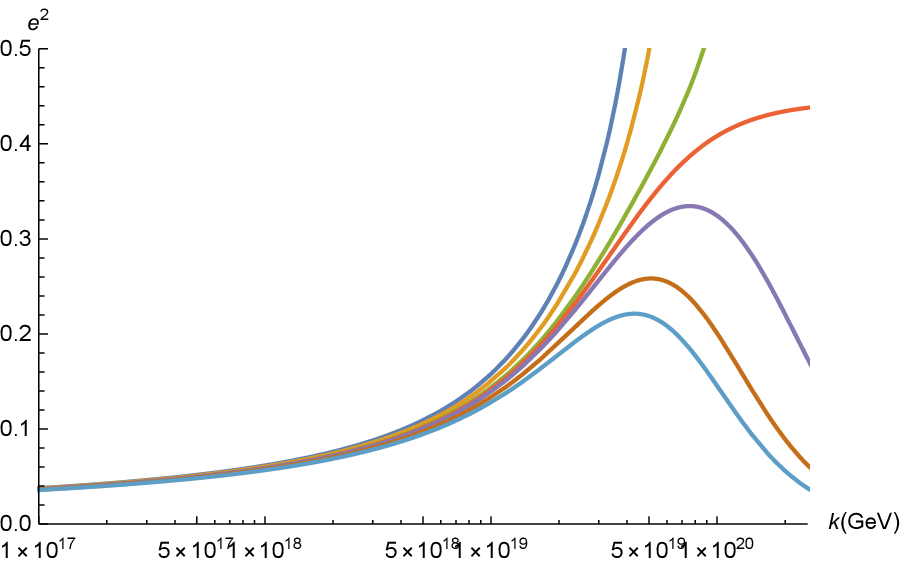}
\end{minipage}
\end{center}
\caption{$e^2(k)$ for decreasing $e_0$ from top to bottom as given in Table~\ref{table4}.
Left for the range of $k=10\sim 3\cdot 10^{20}$ GeV, right for $k=10^{17}\sim 3\cdot 10^{20}$ GeV.
}
\label{fig:U(1)a1}
\end{figure}

We find the corresponding values between $C_0$ and $e_0^2$ in Table~\ref{table4}.
We plot the behaviors of the above coupling (\ref{4.1}) in Fig.~\ref{fig:U(1)a1} for the case of Table~\ref{table4}.
\begin{table}[htb]
\begin{center}
\caption{The corresponding values of $C_0$ and $e_0^2$ for $\til\alpha=1$ and $e_*^2=24\pi^2/533$.}
\vspace{2mm}
\begin{tabular}{|l|l|l|l|l|l|l|l|} \hline
$C_0$   &  $-0.7$ & $-0.4$ & $-0.1$ & 0     & 0.1   & 0.4   & 0.7    \\ \hline
$e_0^2$ &  0.00493  & 0.00492  & 0.00492  & 0.00491 & 0.00491 & 0.00490 & 0.00490 \\ \hline
\end{tabular}\label{table4}
\end{center}
\end{table}

\subsection{Kretschmann scale identification and quantum improved geometries}
\label{subsec:Kretschmann}

Now that we have obtained the running couplings $G(k)$ and $e(k)$, we next relate the cutoff scale $k$
to the radial distance $r$ from the center, thereby obtaining the position-dependent coupling constants $G(r)$
and $e(r)$. The key step is to choose an appropriate scale identification $k(r)$. When we consider
the Schwarzschild metric as our background, any curvature scalar composed of the Ricci tensor cannot be used
to define the scale identification $k(r)$, since the Ricci tensor vanishes identically as a classical vacuum
solution. In this case, the use of the Kretschmann scalar $K := R_{\mu\nu\alpha\beta}R^{\mu\nu\alpha\beta}$
appears to be most appropriate, as it is a diffeomorphism invariant quantity of momentum dimension four,
as discussed in Ref.~\cite{PS18}. The scale identification based on the Kretschmann scalar is given by
\bea
k^4=\chi^4 K(r) \,,
\label{3.14}
\eea
where $\chi$ is a numerical constant. Note that if one employs the quantum Kretschmann scale identification
in which $K$ is computed with respect to the quantum improved metric, the value of $\chi$ may be constrained
as briefly discussed in Ref.~\cite{PS18}.
Note also that when considering an asymptotically de Sitter or AdS spacetime, one may need to subtract
the asymptotic value $K(\infty)$ from (\ref{3.14}).

For the Reissner-Nordstrom metric, the Ricci tensor no longer vanishes. Therefore, we have, in principle,
a larger variety of scale identifications, being able to include those composed of the Ricci tensor and
scalar curvature.
Nevertheless, we employ, as our scale identification $k(r)$,
the above definition \p{3.14} based on the Kretschmann scalar with respect to the classical Reissner-Nordstrom
metric. By doing so, we can directly and clearly compare possible consequences of our case with those of
the quantum improved vacuum black hole cases previously studied in Refs.~\cite{BR00,PS18}.
We will discuss more general choices of scale identification and possible consequences later.

For the classical Reissner-Nordstrom metric~(\ref{def:metric}) with (\ref{metric:RN}), we have
\bea
K_{\rm CRN}(r)=\frac{8G_0^2}{r^{8}}\left(6M^2r^2-12M e_0^2r+7 e_0^4 \right) \,,
\label{3.16}
\eea
where we identify $e_0$ with the low-energy value of the $U(1)$ coupling defined in \p{e0def}.
Inserting \p{3.16} into \p{3.14}, we find the Kretschmann scale identification
\bea
k^2(r) =\frac{2\sqrt{2}\chi^2G_0}{r^4}\sqrt{6M^2r^2-12 M e_0^2 r+7 e_0^4} \,.
\label{3.17}
\label{def:Kretschmann-id}
\eea
We can immediately see the behavior of $k^2(r)$ at short and large distances as
\bea
k^2(r) \simeq \frac{2\sqrt{14} \chi^2 G_0 e_0^2}{r^4} \quad (r \rightarrow 0)\,,
\quad
k^2(r) \simeq \frac{4\sqrt{3}\chi^2 G_0 M}{r^3} \quad (r \rightarrow \infty)\,.
\label{k2:asymptbehave}
\eea

Substituting \p{3.17} into the running couplings $G(k)$ and $e(k)$ found in the previous subsection,
we have the position-dependent couplings $G(r)$ and $e(r)$, and therefore, through \p{3.15},
we obtain the quantum improved lapse function
 \bea
f(r)_{\rm QRN}=1-\frac{2M}{r}G(r)+\frac{G(r) e^2(r)}{ r^2} \,.
\label{3.18}
\eea
To see the behavior of the quantum improved lapse function near the center $r=0$, we note that at short distance,
$r \rightarrow 0$:
\bea
G(r) \simeq A(\til\alpha)r^4 +  O( r^5) \,,  \quad
A(\til\alpha):= \frac{2\pi \til\alpha}{\sqrt{14}\chi^2 G_0 e_0^2} \,.
\label{Gshort}
\eea

\subsubsection{Case $\til{\alpha} \neq1$}

It follows from (\ref{g:UV}) together with (\ref{k2:asymptbehave}) that at short distance, $r \rightarrow 0$:
\bea
 e^2(r) \simeq 4\pi {B}(\til\alpha) {r^{4\til\alpha}} + O(r^{4\til\alpha+1} ) \,,
\label{3.20}
\eea
where the constant $B(\til\alpha)$ is
\bea
\frac{1}{B(\til\alpha)} := 4\pi \left( \frac{\sqrt{14}\chi^2G_0^2 e_0^2}{2\pi \til\alpha} \right)^{\til\alpha}C \,,
\eea
with $C$ given by \p{e0def}.

Inserting \p{Gshort} and \p{3.20} into \p{3.18}, we find that the quantum improved lapse function $f(r)_{\rm QRN}$
behaves near the center as,
\bea
f(r)_{\rm QRN} \simeq 1 - 2MA(\til\alpha)r^3 + {A(\til\alpha)B(\til\alpha)} r^{4\til\alpha +2} \,.
\label{3.22}
\eea
Since $\til \alpha>0$, not only the function $f(r)_{\rm QRN}$ but also the geometry itself is regular
at the center with all curvature scalars vanishing there, as we will discuss later.

At large distances, $r \rightarrow \infty$, it follows from \p{g:IR},
\bea
 e^2 \simeq \frac{ {e_*^2} }{{\til\alpha}\log(1/Dk^2)} \,.
\eea
Thus, together with \p{G:asymptbehave} and \p{k2:asymptbehave}, the large distance behavior is expressed as
\bea
 f(r)_{\rm QRN} \simeq 1 - \frac{2MG_0}{r} + \frac{G_0 e_*^2}{r^2} \cdot \frac{1}{\til\alpha
\log(\pi \til\alpha r^3/\sqrt{3}\chi^2G^2_0 M)} \,.
 \label{asympt:fQRN}
\eea
This renders the metric asymptotically flat in the sense that the resultant quantum improved geometry
approaches the Schwarzschild spacetime at large distances.
However, the geometry does not asymptotically approach the Reissner-Nordstrom spacetime since,
with the logarithmic dependence, the Coulomb potential term decays slightly faster than its classical counterpart.

\begin{figure}[htb]
\begin{center}
\includegraphics[width=70mm]{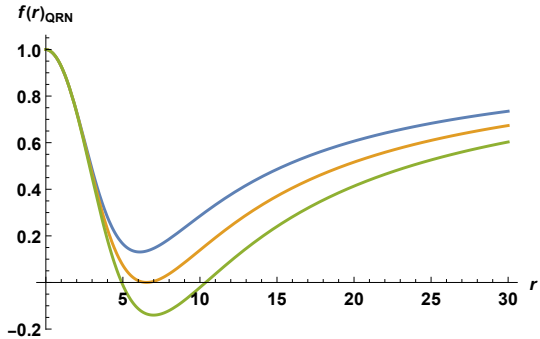}
\end{center}
\caption{$f(r)_{\rm QRN}$ in the Planck unit $G_0=1$ and with $\chi=1$, $\til\alpha =1/52$,
$e_*^2=24\pi^2/533,e_0^2=0.164\,(C=0.1)$, for the increasing mass $M$ from top to bottom:
$M=4$ (blue: no horizon), $4.93$ (orange:  extremal horizon), $6$ (green: two horizons). }
\label{fig:lapse}
\end{figure}

\begin{figure}[htb]
\begin{center}
\includegraphics[width=70mm]{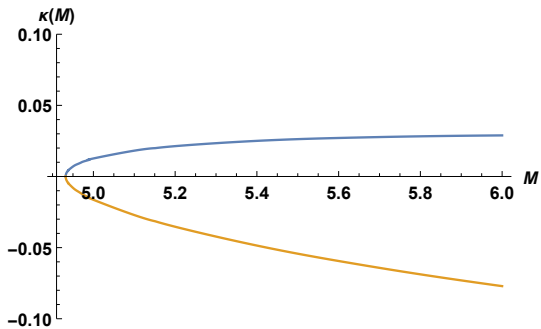}
\end{center}
\caption{The surface gravity as a function of the mass $M$ for the quantum improved Reissner-Nordstrom black hole
with the same parameter values as those in Fig.~\ref{fig:lapse}: $G_0=1$, $\chi=1$, $\til\alpha = 1/52$, $e_*^2 = 24\pi^2/533, e_0^2=0.164 (C=0.1)$.
The upper (blue) curve corresponds to the surface gravity of the event horizon and the lower (orange) curve
to minus the surface gravity of the inner Cauchy horizon. The absolute value of the surface gravity of the Cauchy horizon is larger than that of the even horizon.}
\label{fig:mi}
\end{figure}

Finally, one can check that as in the classical counterpart, the quantum improved Reissner-Nordstorm black hole
can also have, either two horizons (the event and inner Cauchy horizons), one degenerate horizon,
or no horizon depending upon the parameter values. This can be seen in the plot of the quantum improved
lapse function depicted in Fig.~\ref{fig:lapse}. We can also find that for some cases of interest,
the magnitude of the surface gravity of the inner Cauchy horizon is larger than that of the event horizon,
as can be read off from Fig.~\ref{fig:mi}.
This implies that the mass inflation~\cite{PI89,PI90} can also occur for quantum improved Reissner-Nordstrom
black hole, rendering its Cauchy horizon unstable.

\subsubsection{Case $\til\alpha=1$}

Let us turn to the exceptional case $\til\alpha=1$. At short distance, it follows from \p{G:asymptbehave}
and \p{4.2}, $G(k) \simeq {4\pi}/{k^2}$ and $e^2(k) \simeq 4\pi /C_0 G_0k^2$.
Together with (\ref{k2:asymptbehave}), one has
\bea
 G(r) \simeq \frac{2 \pi}{\sqrt{14}\chi^2 G_0 e_0^2} r^4 \,, \quad e^2(r) \simeq \frac{2 \pi }{\sqrt{14}\chi^2 G_0^2 C_0 e_0^2} r^4 \,.
\eea
Therefore at short distance, one obtains,
\bea
f(r)_{\rm QRN}= 1 - \frac{4 \pi M}{\sqrt{14}\chi^2G_0 e_0^2}r^3 + \frac{2 \pi^2}{7 \chi^4 G_0^3 C_0 e_0^4} r^{6} \,.
\eea
Thus, again the quantum improved geometry itself is regular at the center $r=0$.

At large distances, one can find that $f(r)_{\rm QRN}$ behaves the same way as \p{asympt:fQRN} with $\til\alpha\neq 1$.
Again, it is asymptotically flat, but is not precisely the same way as the classical Reissner-Nordstrom
black hole, due to the logarithmic dependence.

\begin{figure}[htb]
\begin{center}
\includegraphics[width=70mm]{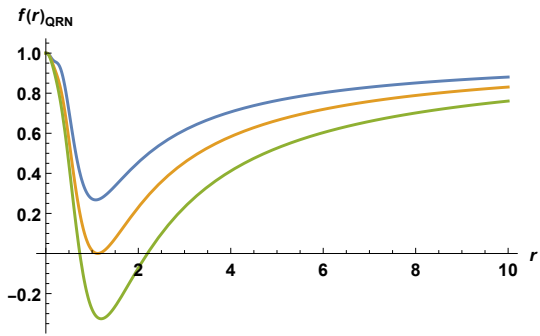}
\end{center}
\caption{$f(r)_{\rm QRN}$ in the Planck unit $G_0=1$ with $\chi=1$, $e_*^2=24\pi^2/533, e_0^2=0.00491$,
for the increasing mass $M$ from top to bottom: $M=0.6$ (blue: no horizon), $0.85$ (orange: extremal horizon),
$1.2$ (green: two horizons).}
\label{fig:lapsea=1}
\end{figure}

We plot the lapse function in Fig.~\ref{fig:lapsea=1} for $\chi=1$, $e_*^2=24\pi^2/533, e_0^2=0.00491$
and various masses. Depending on the parameter values,
the quantum improved metric admits either 2 horizons, 1 degenerate horizon, or no horizon.
We see that the qualitative behavior is similar to the $\til\alpha \neq 1$ case.

\subsection{More general scale identifications and the UV limit}

So far, we have restricted our attention to the scale identification (\ref{3.14}) based on the Kretschmann
scalar~\p{3.16} with respect to the classical metric, and found that our quantum improved geometry is regular
at the center. It is natural to ask what happens, in particular, near the center, if a different scale
identification scheme is chosen.
Here we discuss possible quantum improved black hole geometries in more general setting.

Let us first briefly recapitulate some basic properties of the curvature tensor for static, spherically
symmetric geometry. Consider the general form of the lapse function,
\bea
f(r)=1- \lambda r^{\nu} \,,
\label{lapse:lambda}
\eea
where $\lambda$ and $\nu (\neq 0)$ are some constants. It is immediate to find that the scalar curvature and
the Kretschmann scalar become
\bea
R=\lambda (\nu+1)(\nu+2)r^{\nu-2} \ , \ K= \lambda^2 (\nu^4-2\nu^3+5\nu^2+4)r^{2\nu-4} \,.
\label{RK}
\eea
\vs{-1}
Thus, if $\nu > 2$, those curvature scalars vanish at the center.
In fact, when $\nu > 2$, not only these curvature scalars but all components of the Riemann tensor vanish
at the center. This can easily be seen by introducing the standard orthonormal frame
$\{ e^{(\alpha)}\} = \{ -{f(r)}^{1/2}dt, {f(r)}^{-1/2}dr, rd\theta, r\sin \theta d\phi \}$ and checking that
all components of the curvature tensor with respect to the basis $\{e^{(\alpha)}\}$ are written as
\bea
\label{curvature}
R_{(\alpha)(\beta)(\gamma)(\delta)} = c r^{\nu -2 } \,,
\eea
with $c$ being some constant.
With all curvature components vanishing at the center, the case with $\nu > 2$ can be said to possess
the {\em Minkowski-core}. The marginal case $\nu=2$ can be said to admit the {\em de Sitter(dS)-core}
if $\lambda >0$ and the {\em anti-de Sitter(AdS)-core} if $\lambda <0$.
When $0<\nu<2$, although the lapse function \p{lapse:lambda} is regular at $r=0$, the geometry itself becomes
singular at the center. In the case $0<\nu <2$, however, the central singularity is less divergent than
those appearing in the cases $\nu <0$ such as the Schwarzschild and Reissner-Nordstrom metrics.
In this sense, the case $0< \nu < 2$ may be said to be {\em weakly singular} at the center.

Now let us assume that our scale identification \p{2.2} is given by the following power law in the UV regime:
\bea
k \sim \frac{\xi}{r^p} \,,
\label{gen:scale}
\eea
with a positive constant $p$. The classical Reissner-Nordstrom Kretschmann scale
identification~(\ref{def:Kretschmann-id}) corresponds to $p=2$.
For a general $p>0$, one may have to introduce an appropriate dimensionful parameter to justify the scale
identification (\ref{gen:scale}), but here we leave aside such an issue for simplicity.

One may view that quantum improved charged black holes can be characterized by---besides the classical background
parameters $(G_0M, e_0)$---two quantum improving parameters $(p, \til\alpha)$ which specify the scale
identification and the fixed points of the couplings, respectively.
For the scale identification~(\ref{gen:scale}), one can find that the running couplings behave in the UV as
$ G(r) \sim r^{2p}$ from (\ref{G:asymptbehave}) and $ e^2 \sim r^{2 p \til\alpha}$ from (\ref{g:UV}),
and hence find that the corresponding quantum improved lapse function behaves in the UV as,
\bea
f(r) \sim 1 - \lambda_+ r^{2p -1} + \lambda_- r^{2(p {\til\alpha} + p -1)} \,,
\label{lapse:pm}
\eea
where $\lambda_\pm$ are both positive constants. Roughly speaking, $\lambda_+>0$ describes the strength of
quantum corrected gravitational interaction and $\lambda_->0$ that of quantum corrected electromagnetic interaction.
For example, in the Kretschmann scale identification, $\lambda_+=2MA(\til\alpha) ,
\: \lambda_-=A(\til\alpha)B(\til\alpha)$ (see \p{3.22}).

\begin{figure}[htb]
\begin{center}
\includegraphics[width=110mm]{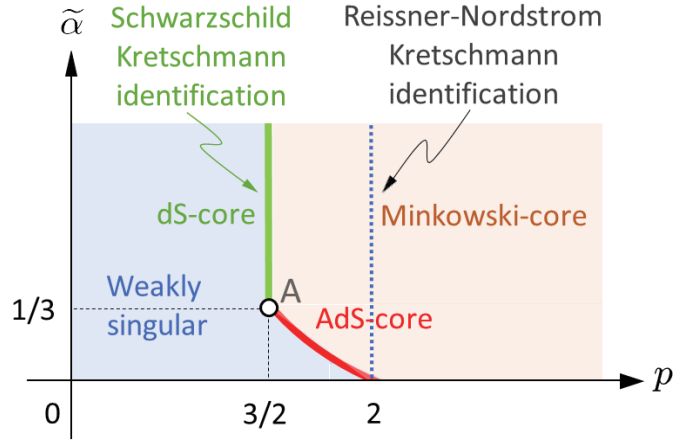}
\end{center}
\caption{The colored (orange) region in the $(p,\til\alpha)$ plane is defined by $p >3/2, \: {\til \alpha} > 2/p-1$,
in which a quantum improved black hole is allowed to have a regular center of Minkowski-core. The lower boundary
(thick red curve) along $\til \alpha = 2/p - 1$ corresponds to AdS-core and the left boundary (green half-line)
along $p=3/2$ to dS-core. At the corner A$(3/2,1/3)$, one has either de Sitter-core, AdS-core, or Minkowski-core,
depending upon the parameter values.
The rest (light-blue) corresponds to weakly singular geometries.
The dashed blue interval $(p=2, 0<\til\alpha )$ corresponds to the case analyzed under the Kretschmann scale
identification in the previous subsection. }
\label{fig:prmtrs}
\end{figure}

Comparing (\ref{lapse:pm}) with (\ref{lapse:lambda}) and inspecting (\ref{RK}) or (\ref{curvature}),
one can find that the regularity at the center is guaranteed when $2p-1 \geq 2, \; 2(p {\til\alpha} + p -1) \geq 2$.
More precisely, one can classify the quantum improved geometries as follows (see also Fig.~\ref{fig:prmtrs}.):
\begin{itemize}
\item $p > \displaystyle\frac{3}{2}, \: \til\alpha > \frac{2}{p}-1$: Minkowski-core.
\item $p = \displaystyle\frac{3}{2}, \: \til\alpha > \frac{1}{3}$:    dS-core.
\item $p > \displaystyle\frac{3}{2}, \: \til\alpha = \frac{2}{p}-1$: AdS-core.
\item $p = \displaystyle\frac{3}{2}, \: \til\alpha = \frac{1}{3}$:    either Minkowski-, dS-, or AdS-core.
\item $0<p<\displaystyle\frac{3}{2}, \:\til\alpha >0$: weak singularity.
\end{itemize}
Note that at the corner $p=3/2, \: {\til\alpha} =1/3$, the geometry admits dS-core for
$\lambda_+ > \lambda_-$, AdS-core for $\lambda_+<\lambda_-$ or Minkowski-core for $\lambda_+ = \lambda_-$.
The dS-core appears only when $p=3/2$, which corresponds to the power of the scale identification with
the radial proper distance, as well as the Kretschmann scalar with respect to the classical Schwarzschild metric,
as shown in Table~\ref{table1}, while the AdS-core appears for all higher power $3/2<p< 2$. The appearance
of the AdS-core is a characteristic property when the running $U(1)$ coupling is included.
The geometries become weakly singular whenever $0<p<3/2$. This can be the case, for example, when $d(r)$
of \p{2.2} is identified with the luminosity distance or the affine parameter of the ingoing null geodesic,
for which $p=1$.
Note that by ``weak singularity" we mean that the Kretschmann scalar of the quantum improved metric is
less divergent than that of the classical counterpart in the short distance limit. This can easily be seen
by comparing (\ref{lapse:pm}) with (\ref{metric:RN}) and
by noting that $2p-1>-1, \: 2(p \til\alpha + p -1)>-2$, provided $p>0, \til\alpha>0$.

\section{Summary and discussions}
\label{sec:4}

In this paper, we have studied the quantum improvement of the charged spherically symmetric black hole
spacetimes in the asymptotic safety scenario. We have first derived the running gravitational and electromagnetic
couplings from the renormalization group equations for the vanishing cosmological constant, and then obtained
the position-dependent running couplings by introducing the scale identification based on the Kretschmann scalar.
Replacing the classical coupling constants in the metric function with the position-dependent running
counterparts, we have constructed the quantum improved Reissner-Nordstrom black hole, which involves
the parameters $\til \alpha$ and $e_*$ that characterize the fixed points of the running gravitational
and $U(1)$ gauge couplings. The global structure of our quantum improved Reissner-Nordstrom black hole is
more or less the same as the classical counterpart, but in the UV limit, the central singularity is
fully resolved, being replaced with a regular Minkowski type core, thanks to the combined effects of
both quantum gravitational and electromagnetic interactions.
The number of the horizons in the quantum improved geometry can be either two (event and Cauchy horizons)
at a maximum, one (degenerate extremal horizon), or zero (no horizon). As for the IR limit, the geometry
at large distances is asymptotically flat but not the same way as its classical counterpart.

Our analysis has been made for the case in which the cosmological constant $\Lambda$ is zero (which is a fixed point).
One might expect that cosmological constant would not significantly affect the short distance behaviors. However,
as shown in~\cite{PS18} for the quantum-improved Schwarzschild-(A)dS case, the quantum improvement
with nonvanishing cosmological constant does not fully remove the central singularity,
in contrast to our case as well as the quantum-improved Schwarzschild case~\cite{BR00}.
So this could be even an interesting criterion to select particle content of the theory.
Our case with vanishing $\Lambda$ has also
the advantage to allow us to find the analytic solution to the renormalization group equation,
making the analysis easier and clearer.

The results should be compared with the pure gravity case, i.e., the quantum improved Schwarzschild black hole,
for which the UV geometry admits the dS-core, but does not the Minkowski-core or AdS-core, under
the Kretschmann scale identification or the scale identification based on the proper radial length.
Our results imply whether or not a classical curvature singularity is resolved and what type of the center
core appears depends upon what types of interactions are involved. It would be interesting to generalize
the present analysis to include Yang-Mills gauge fields, as well as matter fields coupled to the gauge fields.

In this paper we have employed the Kretschmann scale identification as a physically sensible scale identification
scheme.
It is natural to ask how the singularity resolution depends on the type of scale identification.
As briefly mentioned in Sec.~\ref{sec:2}, there are several different scale identification schemes,
which are classified into two types: (i) one based on the proper radial distance of a certain curve of either
timelike or spacelike (or a null curve for which the luminosity distance can be considered) that approaches
the central singularity, and (ii) the other based on the invariant curvature scalars. The former
(i) might be interpreted as an observer dependent scheme by viewing the selected curve
as the world-line of some observer approaching the singularity. For the latter (ii),
one can consider a number of different combinations of the invariant curvature scalars,
$$
\{R, R^\alpha{}_{\beta}R^{\beta}{}_{\alpha},
R^\alpha{}_\beta R^\beta{}_\gamma R^\gamma{}_\alpha,
R^\alpha{}_\beta R^\beta{}_\gamma R^\gamma{}_\delta R^\delta{}_\alpha,
R^{\alpha \beta}{}_{\gamma \delta}R^{\gamma \delta}{}_{\alpha \beta},
R^{\alpha \beta}{}_{\gamma \delta}R^{\gamma \delta}{}_{\rho \sigma}R^{\rho \sigma}{}_{\alpha \beta}, \dots\} \,.
$$
For more detailed list of invariant curvature scalars, see, e.g., Appendix A of \cite{Held21}.
The Kretschmann scalar, $K=R_{\mu \nu \alpha \beta}R^{\mu \nu \alpha \beta}$, is the simplest choice of (ii),
which can apply to the case of vacuum ($R_{\alpha \beta}=0$) classical geometries.
One can employ more complicated curvature invariants
such as those composed of three curvature tensors.
It is clear that the radial dependence $k(r)$ of the cutoff momentum depends on the choice of scale identification.

As a simple generalization, we have examined the case in which the cutoff momentum scale has the inverse power
radial dependence $k \sim 1/r^p$ with $p>0$. We have surveyed the two-dimensional parameter space $(p, \til\alpha)$ and
found that in the $(p, \til\alpha)$-plane, the quantum improved geometries with either dS-core or AdS-core appears
at the boundary between the region for regular geometries with Minkowski-core and that for weakly singular geometries.
The dS-core is formed when the scale identification is made with $p=3/2$, which corresponds to the power of
the classical Schwarzschild Kretschmann scalar.
However, the dS-core is possible only when $1/3< \til\alpha$.
In contrast, the AdS-core can occur for $3/2<p<2$ and $\til\alpha <1/3$, including the value $\til\alpha =1/52$.
It should also be emphasized that the AdS-core is possible
only when the $U(1)$ gauge coupling is involved.

Now to see more on the choice of scale identification and its physical consequences,
let us consider the metric function, (\ref{lapse:pm}), with negative $\nu<0$ so that near the center
$f(r) \sim r^{-|\nu|}$ as in the classical Schwarzschild $\nu=-1$ and Reissner-Nordstrom $\nu=-2$ case.
The proper radial distance toward the singularity is given as $\sim r^{(|\nu|+2)/2}$
(compare Table.~\ref{table1})
and the luminosity distance $\sim r$ for null geodesic, whereas the curvature invariant
$K_n = R^{\alpha \beta}{}_{\gamma \delta} \cdots R^{\rho \sigma}{}_{\alpha \beta}$ that
consists of $n$ curvature tensors behaves typically as $\sim r^{-n(|\nu|+2)}$ near the center\footnote{
This is not always the case when less symmetric geometry is considered,
due to the Lorentzian signature of spacetime metric.}
(see (\ref{curvature})).
Therefore, starting even from the same classical geometry, if a different scale identification is employed,
the resultant quantum improved geometry can differ. For example, the quantum improved Reissner-Nordstrom
geometry has a regular center with $p=2$
for the Kretschmann and proper distance scale identification,\footnote{
This appears to be always the case when one sets
$k=\xi/r^p \propto \{ \mbox{proper radial distance }\}^{-1} \propto (K_n)^{1/2n}$}
while it has a weak singularity with $p=1$ for the scale identification with the luminosity distance along
a null geodesic curve.
In particular, based on curvature invariants, it appears always possible to find a scale identification that attains
a sufficiently large $p$ required for the singularity resolution, provided that a suitable dimensionful
parameter is available. However, apart from the singularity resolution, there seems to
be---at least within the scope of the present paper---no persuasive reason to discard scale identifications
that leads to a weak singularity.
In this regard, it should also be noted that although resolving singularity is an attractive feature,
if all types of singularities are to be resolved, then such a quantum gravity theory may suffer from
pathological features.
For example, the present scheme for quantum improvement can straightforwardly be applied to the negative
mass Schwarzschild solution.
If the singularity of the negative mass Schwarzschild solution is resolved, then admitting a regular negative
energy solution, such a quantum gravity theory would not allow any stable lowest energy solution~\cite{HM95}.
Further analyses are needed to clarify the relation between possible choice of scale identifications
and underlying physics.

The appearance of the AdS core suggests the tantalizing possibility that the present results may be incorporated
into the scenario recently proposed by Ref.~\cite{AEHPAV19}, in which the asymptotic safe scaling regime
can be connected to string theory in the deep UV, where AdS background is more naturally realized.
It would be interesting to pursue this direction further.

\section*{Acknowledgements}
We thank the Yukawa Institute for Theoretical Physis, at Kyoto University, where part of this work was carried
out during the YITP-W-20-09 workshop on ``10th International Conference on Exact Renormalization Group 2020
(ERG2020).'' We would also like to thank Chiang-Mei Chen for valuable discussions.
This work was supported in part by JSPS KAKENHI Grants No. 20K03938, 20K03975, 15K05092 (A. I.),
and in part by 16K05331, 20K03980, and MOST 110-2811-M-008-510 (N.O.).

\bigskip

\noindent
{\it Note added.} After this paper appeared in arXiv, it was pointed out that
there is a closely related paper~\cite{GonzalezKoch2016} which discusses RG improvement of
RN-(a)dS black hole. The main focus of the paper~\cite{GonzalezKoch2016} is on the solution
with nonvanishing cosmological constant, resulting in no resolution of the central singularity.

\end{document}